\documentclass[12pt,aps,showpacs,eqsecnum,nofootinbib,floatfix]{revtex4}
\usepackage{CJK}
\usepackage{bbding}
\usepackage{pifont}
\usepackage{subfigure}
\usepackage{epsfig}
\usepackage{array}
\usepackage{amsmath}
\usepackage{amsfonts}
\usepackage{amssymb}
\usepackage{color}
\usepackage{graphicx}
\usepackage{mathrsfs}
\usepackage{multirow}
\usepackage{caption}

\def\be{\begin{equation}}
\def\ee{\end{equation}}
\def\bea{\begin{eqnarray}}
\def\eea{\end{eqnarray}}

\newcommand{\omits}[1]{}

\begin{document}

\title{Continuous phase transition and microstructure of charged AdS black hole with quintessence}
\author{Xiong-Ying Guo$^{a,b}$, Huai-Fan Li$^{a,b}$\footnote{Email: huaifan.li@stu.xjtu.edu.cn; huaifan999@sxdtdx.edu.cn(H.-F. Li)}, Li-Chun Zhang$^{a,b}$, Ren Zhao$^{b}$}

\medskip

\affiliation{\footnotesize$^a$ Department of Physics, Shanxi Datong
University,  Datong 037009, China\\
\footnotesize$^b$ Institute of Theoretical Physics, Shanxi Datong
University, Datong 037009, China}

\begin{abstract}
Previously, the Maxwell equal-area law has been used to discuss the conditions satisfied by the phase transition of charged
AdS black holes with cloud of string and quintessence, and it was concluded that black holes have phase transition similar to that of vdW system.
The phase transition depends on the electric potential of the black hole and is not the one between a large black hole and a small black hole. On the basis of this result, we study the relation between the latent heat of the phase transition and the parameter of dark energy, and use the Landau continuous phase transition theory to discuss the critical phenomenon of the black hole with quintessence and give the critical exponent. By introducing the number density of the black hole molecules, some properties of the microstructure of black holes are studied in terms of a phase transition. It is found that the electric charge of the black hole and the normalization parameter related to the density of quintessence field play a key role in phase transition. By constructing the binary fluid model of the black hole molecules, we also discuss the microstructure of charged AdS black holes with a cloud of strings and quintessence.
\end{abstract}

 \maketitle

\section{Introduction}
As an interdisciplinary area of general relativity, quantum mechanics, thermodynamics and statistical
physics, particle physics and string theory, black hole physics plays a very important role in modern physics.
The investigation of the thermal properties and the internal microstructure of black holes has always been one of the topics of interest to theoretical physicists. Recently, based on the study of the thermodynamic properties of black holes, some researchers have adopted Boltzmann's profound idea, i.e., a black hole can change its temperature by
absorbing and emitting matter. An object with a temperature has a microstructure. According to this view, the literature~\cite{Wei2015,Wei2019,Wei20192} proposes that black holes are also composed of valid molecules. They think that just as our rooms are filled with air molecules, the interior of a black hole is filled with a kind of ``black hole molecule''. These black hole molecules, of course, give the number of microscopic states of a black hole in a statistical physical sense, and hence the entropy of a black hole. They believe that this internal entropy is equivalent to the ``Bekenstein-Hawking entropy'' of the black hole horizon, and that the black hole molecules themselves carry microscopic degrees of freedom of the black hole's entropy. Based on this hypothesis, the concept of molecular density of black holes is introduced to obtain the phase transition of black holes. It is given by
\begin{equation}
\label{eq1}
m=\frac{1}{v} = \frac{1}{2l_p^2 r_ + },
\end{equation}
where $l_p $ is the Planck length, $l_p = \sqrt {\hbar G / c^3}$.
Introduced $m\equiv (m_{SBH} - m_{LBH} ) / m_c $ with the aid of the
critical molecules number density $m_c = 1 / (2\sqrt 6Q)$. According to the research, the order parameter has a non-zero value when the system passes through the first-order phase transition of the small black hole and the large one. The value of order parameter gradually decrease as the temperature increases, and it equates to zero at the critical point. To some extent, the microstructure of small black holes and large black holes will converge at the critical point. It is found that the variation of the molecular density of the black hole is the internal cause of the black hole phase transition. By studying the ``black hole molecules'' inside a black hole, their collective behavior can be compared to that of a fluid. In other words, these `black hole molecules' are like ``fluid'' inside the black hole. When a black hole is at a certain temperature, given the pressure of these black hole molecules (the cosmological constant), the black hole can undergo a phase transition, known as a ``big black hole'' and ``small black hole'' phase transition, in which the size of the black hole can change suddenly. For example, a large black hole can suddenly shrink in half its size and become a small black hole.

From the view of mathematics, ones use string theory and supersymmetry theory to explain the microstructure of black holes. From the perspective of thermodynamics, the hypothesis of ``black hole molecules'' and ``space-time atoms'' has been proposed to explore the microscopic behavior of black holes~\cite{Rupp2013}. These two aspects of research have made some progress. The microscopic behavior of the thermal stable AdS black hole was studied completely from the thermodynamic point of view by Ruppeiner thermodynamic geometry method, combining the ``black hole molecule'' hypothesis and the concept of microscopic particle number density in ~\cite{Miao2019,Miao1712,Miao1711,Miao1610}, has given a natural explanation for the microscopic behavior of the black hole. It is pointed out that the interaction between black hole molecules in the black hole is mainly attractive, and the description of the molecular potential of ``black hole'' is proposed for the first time. In addition, based on the proposed description of the molecular potential, the correction of the molecular potential to the equation of state of the black hole is calculated and the rationality of the correction term is analyzed.

Recent studies have shown that the RN-AdS black hole has similar thermodynamic critical characteristics as that for van der Waals(vdW) system. And the cosmological constant is interpreted as the pressure in the thermodynamic system~\cite{Mann1207,Mann1301,Kubiznak1608,Kastor09,Kastor10,Cai1306,Seyed5,Hennigar2017,Dinsmore,Gregory,Kastor19,Wei1904,Simovic,Anabalon,Mbarek, 
Hendi1803,Dehghani,Hendi2017,Hendi1811,Hendi1709,Hendi1702,Hendi1706,Panah1805,Panah1906,Dey,Chaturvedi,Ovgun1710,Dayyani1709,Ma1604,Ma1707,Zou1702,Zou2017,Cheng1603,Banerjee1611,Banerjee1109,Banerjee1203,Bhattacharya2017,Cai1606,Zhang1502,Chen19,Chen17,Sabir19,Jafarzade18,Zeng2017,Ma2018,Ma2016,
Ma1708,Li17C,Li17G,Zhang2014,Zhao2013}. With these issues, the thermodynamic characteristics and critical phenomena of AdS and dS black holes are studied, and the first-order and second-order phase transitions of RN-AdS black holes are obtained~\cite{Miao1610,Mann1207,Wei1209}. However, Schwarzschild de Sitter black holes without charge has no phase transition like vdW system~\cite{Zhang1409}. Therefore, the charge of the black hole plays a key role in the phase transition, and the effect of the black hole charge on the phase transition must be considered in the microstructure theory of the black hole.

In the case of a vdW system, the microstructure of the system changes as the fluid undergoes a liquid-gas phase transition, and thus the physical properties of the system change. It is known from the study of black hole system that when the system meets the requirement of thermodynamic equilibrium stability in isothermal or isobaric process, the $P-V$ curve or $T-S$ curve describing the change of the system is discontinuous, and the system has the latent heat of phase transition when it acroses the two-phase coexistence curve. The system has different physical properties in two phases and thus has different microstructure in different phases. In the literature~\cite{Guo}, Maxwell equal-area law was used to discuss the critical behavior of the RN-AdS black hole, and it was found that the generation of the phase transition of the RN-AdS black hole was mainly determined by the electric potential at the event horizon of the black hole, rather than simply determined by the size of the black hole.
The recent observations, showed that the dark energy is prevailing in our
universe~\cite{Komatsu,Riess}. The equation of state of dark energy approaches to
cosmological constant or vacuum energy~\cite{Peebles}, however, the dynamical dark
energy is also possible~\cite{Copeland}, and the quintessence could affect the black
hole space time~\cite{Stuchl}.

It is showed that, in the presence of the quintessence, the
charged AdS black hole exhibits a small-large black hole phase transition,
which is similar to the liquid-gas phase transition of vdW fluid. Near the
critical point, the heat capacity will diverge. It also shares the same
critical phenomena and scaling laws~\cite{Li14,Wei18C}. For there are the dark energy corresponding state parameter $\alpha$
and state equation $p =\omega\rho$
corresponding parameter $\omega$ in the charged AdS black holes
with cloud of string and quintessence(charged AdSQ black hole). So, the study on critical behavior of charged
AdSQ black hole must discuss the effect of $\alpha$ and $\omega$ on phase transition.
In order to make the conclusion universal, two groups of
independent variables $P-V$ and $T-S$ were selected to discuss the
critical behavior of charged AdSQ black hole. The microscopic interpretation of black hole phase
transition and the critical exponential thermodynamic geometry method, as a unique perspective, played an important
role in the study of black hole phase transition. Ruppeiner's clear physical picture makes it widely used~\cite{Miao1712,Miao1711,Miao1610,Ruppeiner2008,Ruppeiner1995,Seyed1,Seyed2,Seyed3,Seyed4,Mirza2008,Mirza2009}.

This paper in organized as follows. In the section (\ref{cabh}),
we give a briefly review for the thermodynamic quantity of charged AdSQ black hole.
In this section(\ref{eal}), we choose different conjugate variables and use Maxwell's equal-area law to
study the phase transition of charged AdSQ black hole, which is not a simple phase transition between
a small black hole and a large black hole. The first order phase transition of a black
hole is a mutation for the potential $\phi = \frac{Q}{r_ +}$ and quintessence potential $\Theta=\frac{3\alpha}{r_+^{3\omega+1}}$.
On the other hand, this reflects that black holes produce first order liquid-vapor phase
transitions similar to vdW systems and have coexistence zones. Because the charged AdSQ
black hole molecule is affected by electric potential (electric field) and quintessence
potential, the black hole molecule produces orientation polarization and displacement
polarization, which makes the black hole molecule have certain orientation. Under the dual action of quintessence
potential and thermal motion, the orientation degree of the black hole molecules is different.
The orientation degree of the black hole molecules determines the phase of the black hole. Guided by this thought,
in the section (\ref{laten}), one analysis the influence of each parameter on the latent heat of phase
transition when the first-order phase transition occurred in charged AdSQ black hole.
In the section (\ref{micro}), electric potential and quintessence potential were selected as the order parameters for
studying the black hole phase transition, and Landau continuous phase transition theory was used to analyze
the continuous phase transition of the charged AdSQ black hole.
 In the section (\ref{geo}), tries to understand some properties of the microstructure of black hole
 molecules by analyzing the influence of the value of $x$ under different parameters (that is, taking different temperatures) on $R$.
The section (\ref{con}) is a summary. For
simplicity, we adopt in the following the units $\hbar=c=k_B=G=1$ .

\section{charged AdS black hole}
\label{cabh}

In the presence of the quintessence, the line element of a charged AdS black
hole is~\cite{Kiselev,Komatsu,Riess,Peebles,Copeland,Stuchl,Li14,Ma17,Wei18C}
\begin{equation}
\label{eq2}
ds^2 = - f(r)dt^2 + f^{ - 1}dr^2 + r^2d\Omega _2^2 ,
\end{equation}
where the metric function is given by
\begin{equation}
\label{eq3}
f(r) = 1 - \frac{2M}{r} + \frac{Q^2}{r^2} - \frac{\Lambda r^2}{3} - \alpha
r^{ - 3\omega - 1}.
\end{equation}
Here the parameters $M$, $Q$ and $\Lambda $ are, respectively, related to
the black hole mass, charge and cosmological constant. $\omega $ is the
state parameter for quintessence matter under the state equation $p = \omega
\rho$. If $\omega$ satisfies $-1<\omega<-1/3$, the quintessence
will make the universe acceleration expansion. The parameter $\alpha$ is related to
the energy density of quintessence matter.

In the extended phase space, the cosmological constant is treated as
thermodynamic pressure
\begin{equation}
\label{eq4}
P = - \frac{\Lambda }{8\pi }.
\end{equation}
Solving the equation $f(r_+)=0$, one can obtain the event horizon radius $r_+$,
with which the mass of the black hole mass can be
expressed as
\begin{equation}
\label{eq5}
M=\frac{r_ + }{2} + \frac{Q^2}{2r_ + } + \frac{4\pi P}{3}r_ + ^3 -
\frac{\alpha }{2}r_ + ^{ - 3\omega }.
\end{equation}
According to the Bekenstein-Hawking entropy-area relation, the black hole
entropy is
\begin{equation}
\label{eq6}
S = \frac{A}{4} = \pi r_ + ^2 .
\end{equation}
Treating state parameter $\alpha$ as a new thermodynamic quantity, the
first law of the black hole is~\cite{Li14,Wei18C}
\begin{equation}
\label{eq7}
dM = TdS + \phi dQ + VdP + \Theta d\alpha .
\end{equation}
Employing it, we can obtain the thermodynamic quantities
\[
T = \frac{f'(r_ + )}{4\pi } = \frac{1}{4\pi r_ + }\left( {1 - \frac{Q^2}{r_
+ ^2 } + 8\pi Pr_ + ^2 + 3\omega \alpha r_ + ^{ - 3\omega - 1} } \right),
\]
\begin{equation}
\label{eq8}
\phi = \frac{Q}{r_ + },
\quad
V = \frac{4}{3}\pi r_ + ^3 ,
\quad
\Theta = - \frac{1}{2}r_ + ^{ - 3\omega } .
\end{equation}
It also shares the same critical behavior as that of the vdW fluid.
Especially, when the parameter $\omega=-2/3$, the analytical critical
point can be obtained~\cite{Li14}
\begin{equation}
\label{eq9}
r_ + ^c = \sqrt 6 Q,
\quad
T_c = \frac{\sqrt 6 }{18\pi Q} - \frac{\alpha }{2\pi },
\quad
P_c = \frac{1}{96\pi Q^2}.
\end{equation}

\section{The equal-area law of charged AdSQ black hole in extended phase space}
\label{eal}
Keeping $Q$, $\alpha $ and $\omega$ of charged AdSQ black hole as constants, for Eq.(\ref{eq8}),
the equation of states of charged AdSQ black
hole is analogous to the simple system of the general thermodynamics system. The equation of states can
be written as the form of $f(T,P,V)=0$. First, we use Maxwell equal-area law to study
the condition of phase transition occurs with $Q$, $\alpha$
and $\omega$ fixed and with differential conjugate variables $P-V$ and $T-S$.

\subsubsection{The construction of equal-area law in $P-V$ diagram}
\label{subsubsec:mylabel1}

Taking $Q$, $\alpha $ and $\omega$ of charged AdSQ black hole as constants,
the temperature $T_0$ ($T_0 \le T_c )$, $T_c $ is critical temperature. The horizontal axes
of the two-phase coexistence region are $V_2$ and $V_1$, and the vertical axis is $P_0$,
which depends on the radius $r_+$ of the black hole horizon. We can obtain the following
expression by Maxwell's equal area laws
\begin{equation}
\label{eq10}
P_0 (V_2 - V_1 ) = \int\limits_{V_1 }^{V_2 } {PdV}=\int\limits_{r_1 }^{r_2 } {4 P(r)\pi r^2 dr}.
\end{equation}
From Eq.(\ref{eq10}), We can obtain
\begin{equation}
\label{eq11}
P_0 = \frac{T_0 }{2r_1 } - \frac{1}{8\pi r_1^2 } + \frac{Q^2}{8\pi r_1^4 } -
\frac{3\omega \alpha }{8\pi }r_1^{ - 3\omega - 3} ,
\quad
P_0 = \frac{T_0 }{2r_2 } - \frac{1}{8\pi r_2^2 } + \frac{Q^2}{8\pi r_2^4 } -
\frac{3\omega \alpha }{8\pi }r_2^{ - 3\omega - 3},
\end{equation}
\begin{equation}
\label{eq12}
2P_0 = \frac{3T_0 (1 + x)}{2r_2 (1 + x + x^2)} - \frac{3}{4\pi r_2^2 (1 + x
+ x^2)} + \frac{3Q^2}{4\pi r_2^4 x(1 + x + x^2)} - \frac{3\alpha (1 -
x^{3\omega })}{4\pi r_2^{3\omega + 3} x^{3\omega }(1 - x^3)},
\end{equation}
where $x = r_1 / r_2 $. From the Eq.(\ref{eq11}), we can obtain
\begin{equation}
\label{eq13}
0 = T_0 - \frac{1}{4\pi r_2 x}(1 + x) + \frac{Q^2}{4\pi r_2^3 x^3}(1 +
x^2)(1 + x) - \frac{3\omega \alpha (1 - x^{3\omega + 3})}{4\pi r_2^{3\omega
+ 2} x^{3\omega + 2}(1 - x)},
\end{equation}
\begin{equation}
\label{eq14}
2P_0 = \frac{T_0 }{2r_2 x}(1 + x) - \frac{1}{8\pi r_2^2 x^2}(1 + x^2) +
\frac{Q^2}{8\pi r_2^4 x^4}(1 + x^4) - \frac{3\omega \alpha (1 + x^{3\omega +
3})}{8\pi r_2^{3\omega + 3} x^{3\omega + 3}}.
\end{equation}
From Eqs.(\ref{eq12}) and (\ref{eq14}), we can obtain
\[
\frac{1}{4\pi r_2 x} = \frac{T_0 (1 + x)(1 - x)^2}{(1 + x - 4x^2 + x^3 +
x^4)} + \frac{Q^2[(1 + x^4)(1 + x + x^2) - 6x^3]}{4\pi r_2^3 x^3(1 + x -
4x^2 + x^3 + x^4)}
\]
\begin{equation}
\label{eq15}
 - \frac{3\alpha \left[ {\omega (1 + x^{3\omega + 3})(1 - x^3) - 2x^3(1 -
x^{3\omega })} \right]}{4\pi r_2^{3\omega + 2} x^{3\omega + 2}(1 - x)(1 + x
- 4x^2 + x^3 + x^4)}.
\end{equation}
Form Eq.(\ref{eq13}), we obtain
\begin{equation}
\label{eq16}
T_0 = \frac{(1 + x)}{4\pi r_2 x} - \frac{Q^2}{4\pi r_2^3 x^3}(1 + x)(1 +
x^2) + \frac{3\omega \alpha (1 - x^{3\omega + 3})}{4\pi r_2^{3\omega + 2}
x^{3\omega + 2}(1 - x)}.
\end{equation}
Substituting Eq.(\ref{eq16}) into (\ref{eq15}), we get
\[
r_2^2 = \frac{Q^2}{x^2}\frac{(1 + 2x - 6x^2 + 2x^3 + x^4)}{(1 - x)^2}
\]
\[
 + \frac{3\alpha }{r_2^{3\omega - 1} x^{3\omega + 2}(1 - x)^3}\left[ {\omega
(1 - x^{3\omega + 3})(1 - x^2)(1 - x) - \omega (1 + x^{3\omega + 3})(1 -
x^3) + 2x^3(1 - x^{3\omega })} \right]
\]
\begin{equation}
\label{eq17}
 = Q^2f_1 (x) + \frac{3\alpha }{r_2^{3\omega - 1} }f_2 (x,\omega ) = Q^2f_1
(x) - r_2^2\Theta _2 f_2(x,\omega ),
\end{equation}
where $f_1 (x) = \frac{(1 + 4x + x^2)}{x^2}$, the quintessence potential of the black hole horizon $\Theta _2 =
\frac{3\alpha }{r_2^{3\omega + 1} }$,
\[
f_2 (x,\omega ) = \frac{\omega (1 - x^{3\omega + 3})(1 - x^2)(1 - x) -
\omega (1 + x^{3\omega + 3})(1 - x^3) + 2x^3(1 - x^{3\omega })}{x^{3\omega +
2}(1 - x)^3}.
\]
When $\omega = - 2 / 3$, $_{ }f_2 (x, - 2 / 3) = 0$, Eq.(\ref{eq17}) can be written as  $r_2^2 =
\frac{Q^2}{x^2}(1 + 4x + x^2) = Q^2f_1 (x)$, this result identify with the one of RN-AdS black hole~\cite{Guo}.
When $\omega \ne - 2 / 3$, $r_1 = r_2 = r_c$, Form Eq.(\ref{eq17}) we can find that the horizon radius $r_c$
of critical point satisfies
\[
r_c^2 = 6Q^2 - \frac{9\omega \alpha }{2r_c^{3\omega - 1} }(\omega +
1)(3\omega + 2) = 6Q^2 + \frac{3r_c^2 }{4\pi }B_c (\omega + 1)(3\omega + 2),
\]
so,
\begin{equation}
\label{eq19}
\frac{1}{6}=\frac{Q^2}{r_c^2 } + \frac{\rho _c A_c }{8\pi }(\omega +
1)(3\omega + 2) = \phi _c^2 + \frac{B_c }{8\pi }(\omega + 1)(3\omega +
2),
\end{equation}
where $A_c = 4\pi r_c^2 $ is the area of the critical horizons, $\rho _c = - \frac{3\omega
\alpha }{2r_c^{3(\omega + 1)} }$ is density of quintessence of the critical horizons,
$B_c = \rho _c A_c = - \frac{6\omega \alpha \pi }{r_c^{3\omega + 1}}$ is the quintessence
of the unit thickness of the black hole horizon, $\phi _c$ is the charged
potential of the critical horizon, $r_c $ is the horizon radius.

When $B_c$ and $\omega$ is given, from Eq.(\ref{eq19}) we can obtain $\phi_c$.
When the charged $Q$ is given, from $\phi_c$ we can get the critical radius $r_c$.
According to Eq. (\ref{eq19}), when the parameter $\omega$ was set, there are a second phase transition for the charged AdSQ black hole
. The sum $B_c$ of electric potential $\phi _c$ and quintessence on the event horizon of charged AdSQ black hole is a constant.
Therefore, the position of the second-order phase transition point of the black hole is determined by
the electric potential $\phi _c$ at the event horizon of the black hole and the sum of
quintessence potential per unit thickness on the event horizon of the black hole.
When $\omega=-2/3$, the second phase transition position of charged AdSQ black hole is
only related to electric potential, which is the same as charged AdS black hole~\cite{Guo}.

When $x \to 1$, from Eq. (\ref{eq16}) we can obtain the critical temperature $T_c $, which satisfies
\begin{equation}
\label{eq20}
T_c = \frac{1}{3\pi r_c } - \frac{3\omega \alpha (3\omega - 1)(\omega +
1)}{4\pi r_c^{3\omega + 2} } = \frac{1}{3\pi r_c }\left( {1 + \frac{3B_c
(3\omega - 1)(\omega + 1)}{8\pi }} \right).
\end{equation}
When $x \to 1$, from Eq.(\ref{eq12}) we can obtain the critical pressure $P_c $, which satisfies
\begin{equation}
\label{eq21}
P_c = \frac{1}{6\pi r_c^2 } + \frac{Q^2}{8\pi r_c^4 } - \frac{3\omega
^2\alpha (3\omega + 2)}{8\pi r_c^{3\omega + 3} } = \frac{1}{6\pi r_c^2
}\left( {1 + \frac{3}{4}\phi _c^2 + \frac{3\omega B_c (3\omega + 2)}{8\pi
}} \right).
\end{equation}
When $\omega = -2/3 $, critical position $r_c$, critical temperature $T_c$
and critical pressure return to known conclusion~\cite{Chaturvedi,Li14,Wei18C}.
Substituting Eq.(\ref{eq17}) into Eq.(\ref{eq16}), we get
\begin{equation}
\label{eq23}
T_0 = \frac{(1 + x)}{\pi r_2 x^2f_1 (x)} + \frac{3\alpha }{4\pi
xr_2^{3\omega + 2} }\left[ {\frac{(1 + x)(1 + x^2)f_2 (x,\omega )}{1 + 4x +
x^2} + \frac{\omega (1 - x^{3\omega + 3})}{x^{3\omega + 1}(1 - x)}}
\right].
\end{equation}
Taking $T_0 = \chi T_c = \frac{\chi }{3\pi r_c } - \chi \frac{3\omega \alpha
(3\omega - 1)(\omega + 1)}{4\pi r_c^{3\omega + 2} }$, with
$0<\chi\le1$, and substituting it into Eq.(\ref{eq23}), we can obtain
\begin{equation}
\frac{\chi }{3r_c }\left( {1 + \frac{3B_c (3\omega - 1)(\omega + 1)}{8\pi }}
\right) = \frac{(1 + x)}{r_2 x^2f_1 (x)} - \frac{B_c r_c^{3\omega + 1}
}{8\pi \omega xr_2^{3\omega + 2} }\left[ {\frac{(1 + x)(1 + x^2)f_2
(x,\omega )}{1 + 4x + x^2} + \frac{\omega (1 - x^{3\omega + 3})}{x^{3\omega
+ 1}(1 - x)}} \right].
\label{eq231}
\end{equation}
For given $\chi$, $B_c$, $\omega$ and $Q$, substituting Eq. (\ref{eq17}) into Eq.(\ref{eq231}),
we can obtain the value of $x$ at temperature. Substituting the obtained value $x$ into Eq.(\ref{eq17}), we can get $r_2$,
and get $r_1$ from $r_1 = x r_2$.

\begin{figure}[!htbp]
\center{\includegraphics[width=5cm,keepaspectratio]{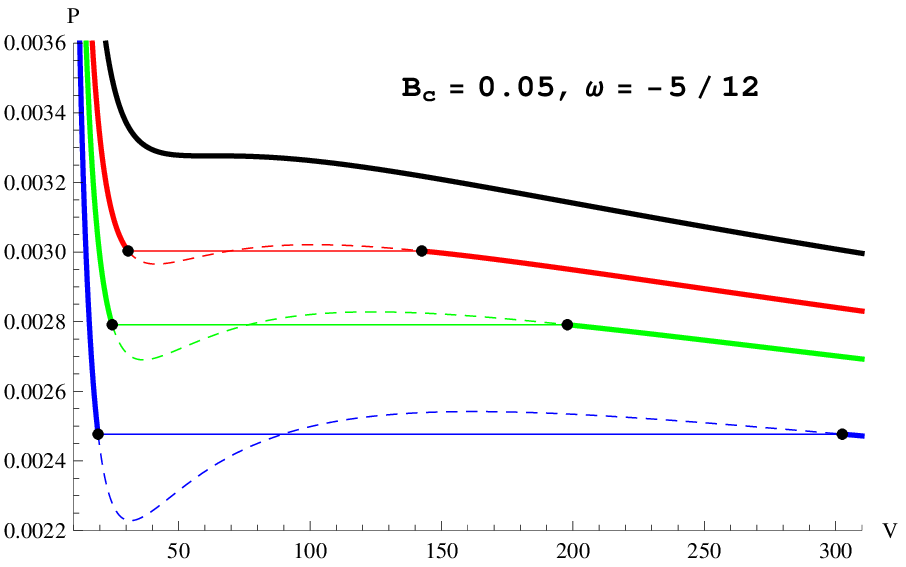}
\includegraphics[width=5cm,keepaspectratio]{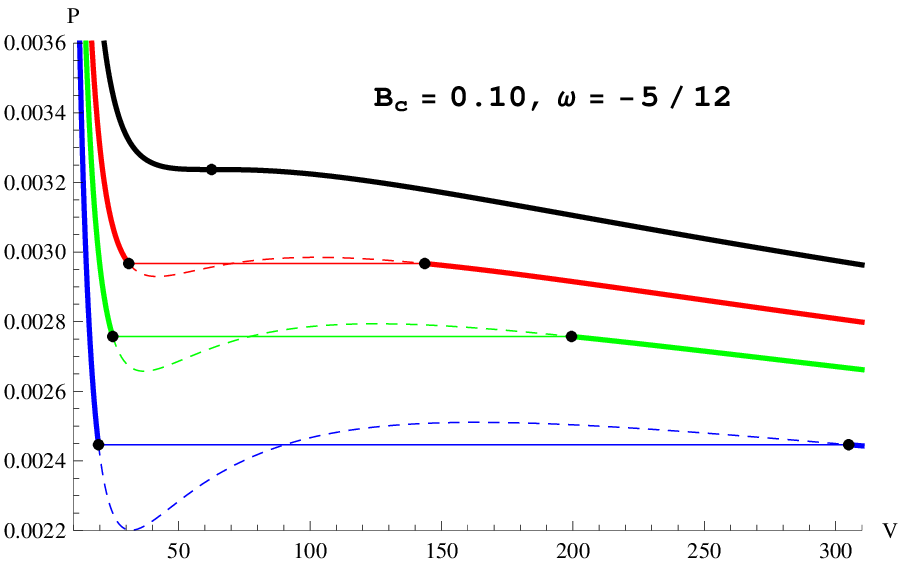}
\includegraphics[width=5cm,keepaspectratio]{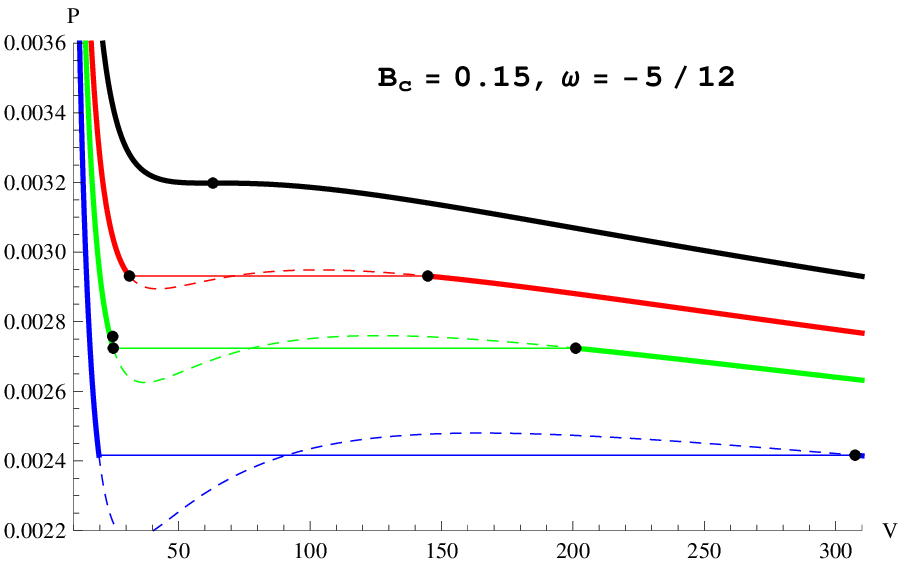}
\includegraphics[width=5cm,keepaspectratio]{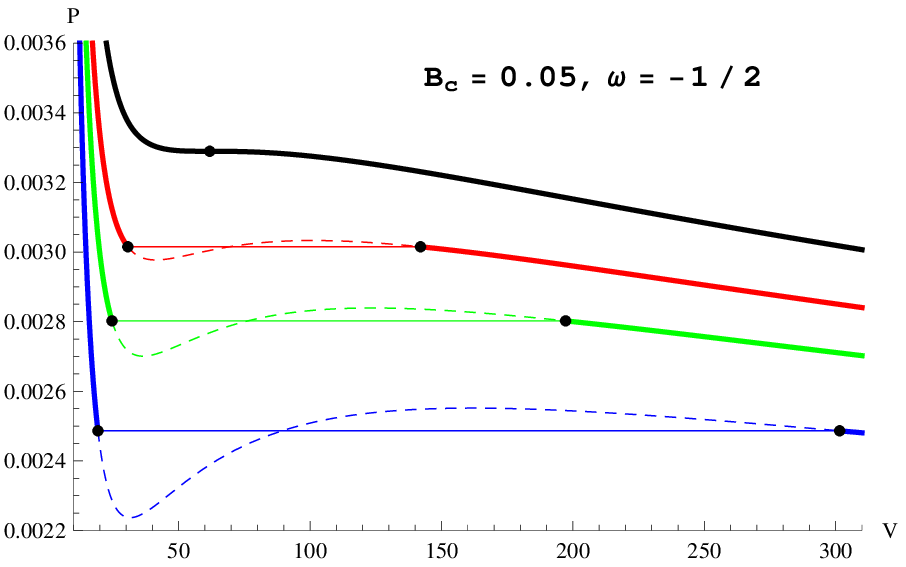}
\includegraphics[width=5cm,keepaspectratio]{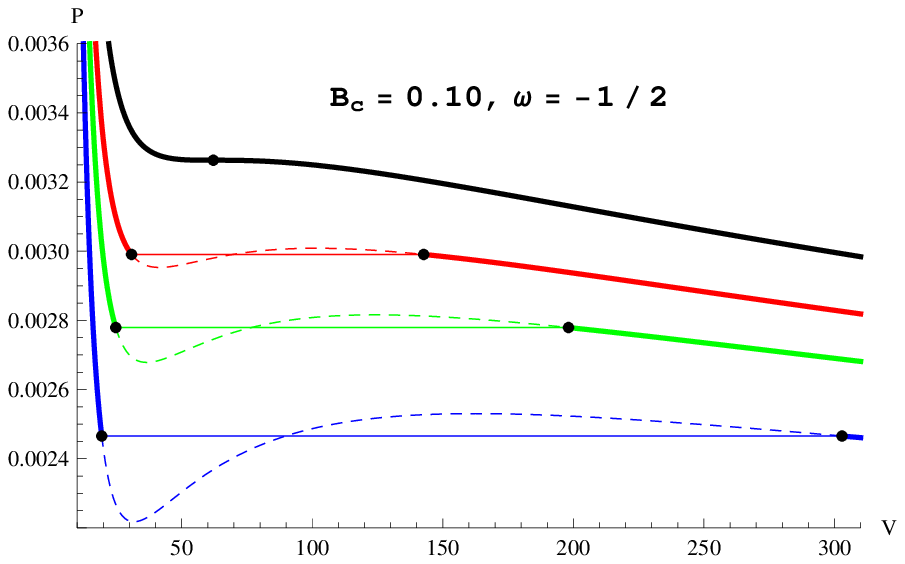}
\includegraphics[width=5cm,keepaspectratio]{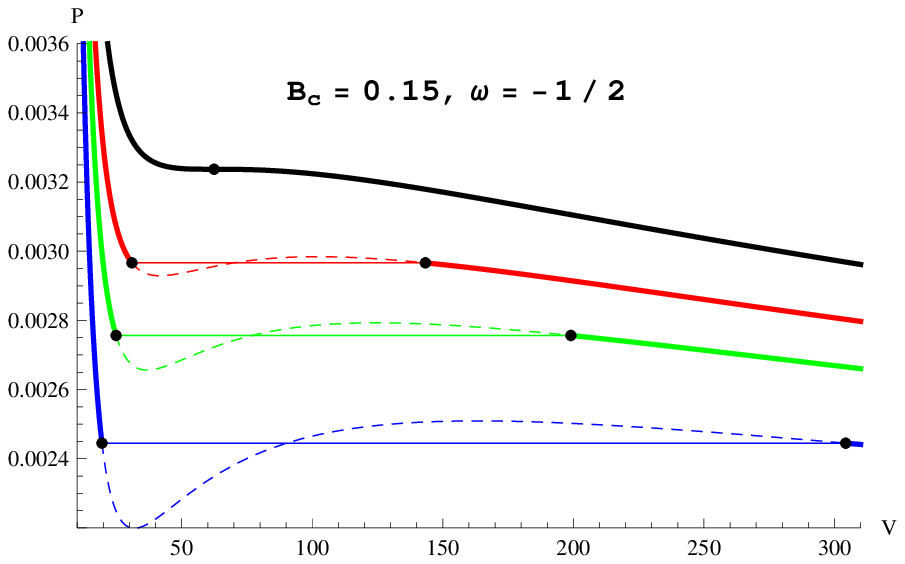}
\includegraphics[width=5cm,keepaspectratio]{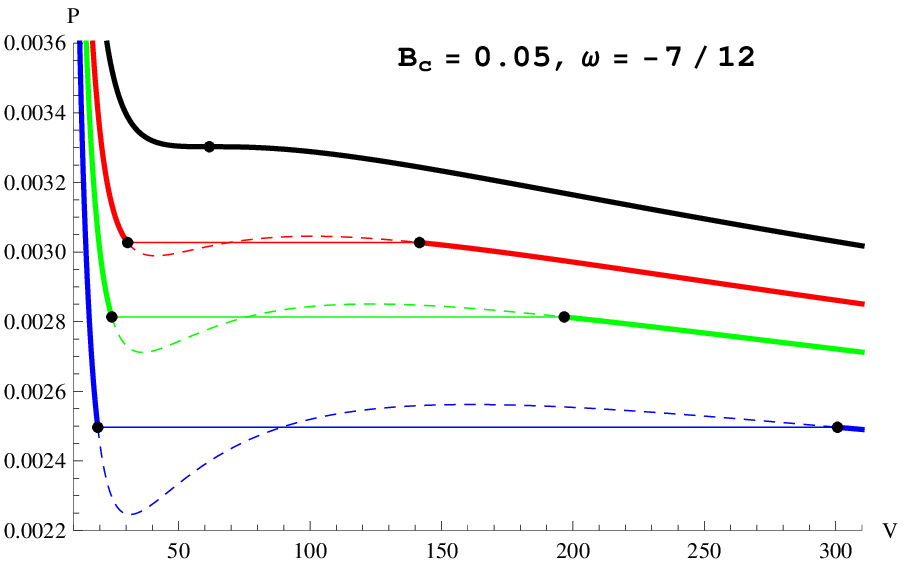}
\includegraphics[width=5cm,keepaspectratio]{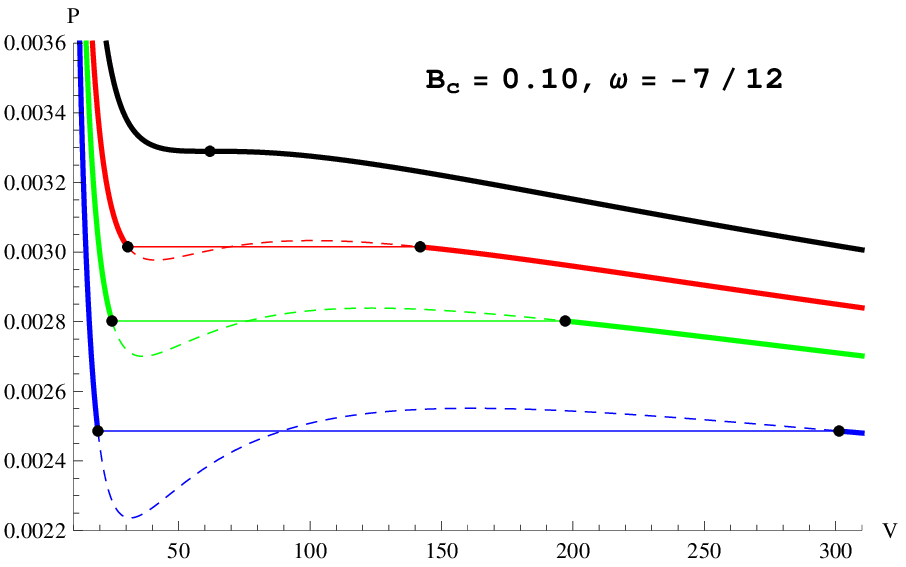}
\includegraphics[width=5cm,keepaspectratio]{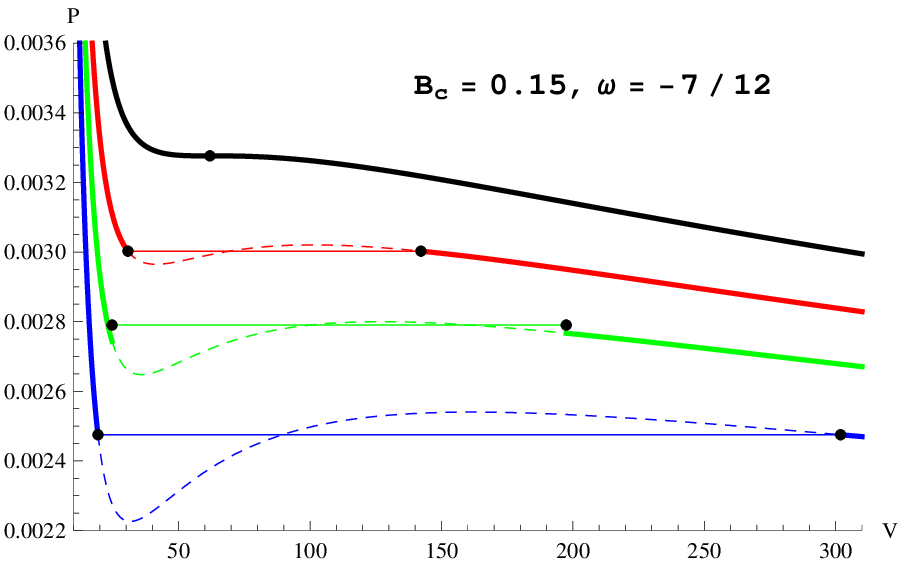}\hspace{0.5cm}}\\
\captionsetup{font={scriptsize}}
\caption{(Color online) The simulated phase transition and the boundary of a two-phase
coexistence on the base of isotherms in the $P-V$ diagram for charged anti-de Sitter black holes.
The temperature of isotherms decreases from top to bottom. The green line is the critical isotherm. \label{PVR}}
\end{figure}

\begin{figure}[!htbp]
\center{\includegraphics[width=8.1cm,keepaspectratio]{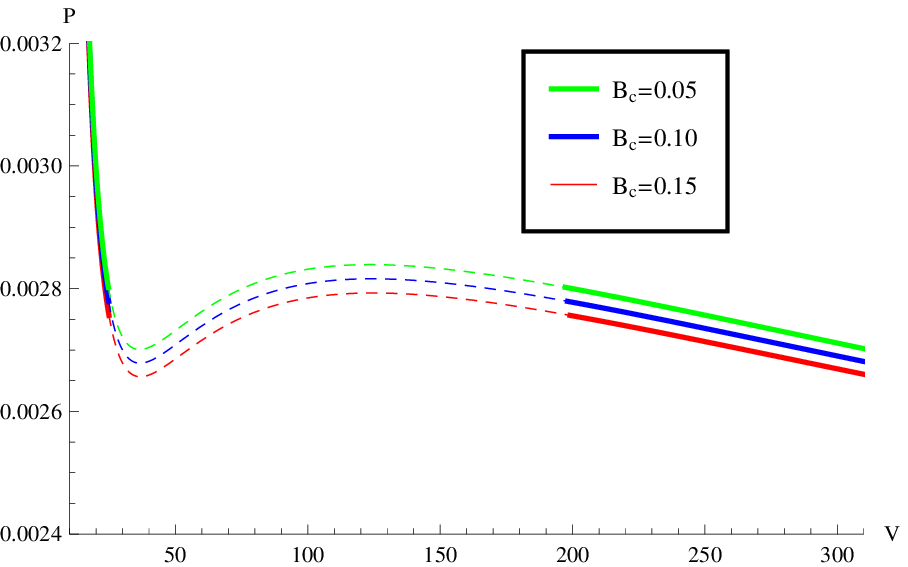}
\includegraphics[width=8.1cm,keepaspectratio]{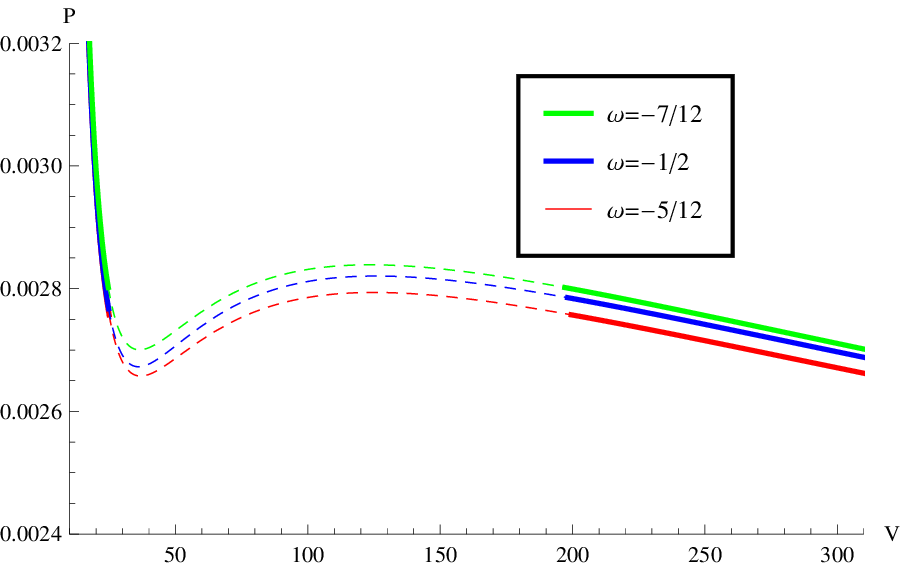}
\hspace{0.5cm}}\\
\captionsetup{font={scriptsize}}
\caption{(Color online) The simulated phase transition and the boundary of a two-phase
coexistence on the base of isotherms in the $P-V$ diagram for charged anti-de Sitter black holes with differential $B_c
$ and $\omega $. The left fig. $Q = 1,\omega = - 1 / 2$, the right fig.$Q = 1,B_c = 0.1$
\label{PVR1}}
\end{figure}

From
\begin{equation}
\label{eq24}
P = \frac{\chi T_c }{2r_ + } - \frac{1}{8\pi r_ + ^2 } + \frac{Q^2}{8\pi r_
+ ^4 } + \frac{B_c r_c^{3\omega + 1} }{16\pi ^2}r_ + ^{ - 3\omega - 3} ,
\end{equation}
we can give the graph $P-V$ for the temperature $T_0 = \chi T_c $, and constant $Q$, $B_c$ and $\omega \ne -2/3$ in Fig.(\ref{PVR}).
The horizontal line in the Fig.(\ref{PVR}) is the region where the two phases coexist, and the intersection point of the horizontal line and the curve is the position of the first-order phase transition point of the black hole. To show the effect of the parameters $B_c$ and $\omega$ on the phase transition point, at the same temperature, we draw the $P - V$ curves of different $B_c$ and $\omega$ in Fig.(\ref{PVR1}).
As shown in Fig.(\ref{PVR1}), the pressure for the phase transition point increase, and the coexistence zone of the two phases decreases with the value $B_c$ of decrease at the same black hole temperature. The pressure for the phase transition point increase and the coexistence zone of the two phases decreases as the value of $\omega$ decreases at the same black hole temperature.

\subsubsection{The construction of equal-area law in $T-S$ diagram}

Taking the invariable cosmological constant $l$, the horizontal axes of the
two-phase coexistence region are $S_2$ and $S_1$, respectively. The vertical axis
is $T_0 (T_0 \le T_c )$ which depends on the radius $r_+$ of the black hole event horizon. We can
obtain by Maxwell's equal area laws
\begin{equation}
\label{eq25}
T_0 (S_2 - S_1 ) = \int\limits_{S_1 }^{S_2 } {TdS_ + } = \int\limits_{r_1
}^{r_2 } {\frac{1}{2}\left( {1 + \frac{3r_ + ^2 }{l^2} - \frac{Q^2}{r_ + ^2
} + 3\omega \alpha r_ + ^{ - 3\omega - 1} } \right)dr_ + } ,
\end{equation}
\begin{equation}
\label{eq26}
T_0 = \frac{1}{4\pi r_2 }\left( {1 + \frac{3r_2^2 }{l^2} - \frac{Q^2}{r_2^2
} + 3\omega \alpha r_2^{ - 3\omega - 1} } \right),
\quad
T_0 = \frac{1}{4\pi r_1 }\left( {1 + \frac{3r_1^2 }{l^2} - \frac{Q^2}{r_1^2
} + 3\omega \alpha r_1^{ - 3\omega - 1} } \right).
\end{equation}
From Eq.(\ref{eq25}) we can obtain
\begin{equation}
\label{eq27}
2\pi T_0 r_2^2 (1 - x^2) = r_2 (1 - x) + \frac{r_2^3 }{l^2}(1 - x^3) -
Q^2\frac{1 - x}{r_2 x} + \frac{\alpha (1 - x^{3\omega })}{r_2^{3\omega }
x^{3\omega }}.
\end{equation}
From Eq.(\ref{eq26}) we get
\begin{equation}
\label{eq28}
0 = - \frac{1 - x}{r_2 x} + \frac{3r_2 }{l^2}(1 - x) + \frac{Q^2}{r_2^3
x^3}(1 - x^3) - \frac{3\omega \alpha (1 - x^{3\omega + 2})}{r_2^{3\omega +
2} x^{3\omega + 2}},
\end{equation}
\begin{equation}
\label{eq29}
8\pi T_0 = \frac{1 + x}{r_2 x} + \frac{3}{l^2}r_2 (1 + x) - \frac{Q^2}{r_2^3
x^3}(1 + x^3) + \frac{3\omega \alpha (1 + x^{3\omega + 2})}{r_2^{3\omega +
2} x^{3\omega + 2}}.
\end{equation}
From Eqs.(\ref{eq27}), (\ref{eq28}) and (\ref{eq29}) we can obtain
\begin{equation}
\label{eq291}
\frac{r_2^2 }{l^2}x = 1 - \frac{Q^2(1 + x - 4x^2 + x^3 + x^4)}{r_2^2 x^2(1 -
x)^2} + \frac{3\omega \alpha (1 + x^{3\omega + 2})(1 - x^2) - 4\alpha x^2(1
- x^{3\omega })}{r_2^{3\omega + 1} x^{3\omega + 1}(1 - x)^3}.
\end{equation}
Substituting Eq.(\ref{eq28}) into Eq.(\ref{eq291}), we get
\begin{equation}
\label{eq30}
1 = \frac{Q^2(1 + 2x - 6x^2 + 2x^3 + x^4)}{r_2^2 x^2(1 - x^2)} -
\frac{3\alpha [\omega (1 - x)(1 + 2x + 2x^{3\omega + 2} + x^{3\omega + 3}) -
2x^2(1 - x^{3\omega })]}{r_2^{3\omega + 1} x^{3\omega + 1}(1 - x)^3},
\end{equation}
we find that Eq. (\ref{eq30}) is equivalent to Eq. (\ref{eq17}).

\begin{figure}[!htbp]
\center{\includegraphics[width=8.1cm,keepaspectratio]{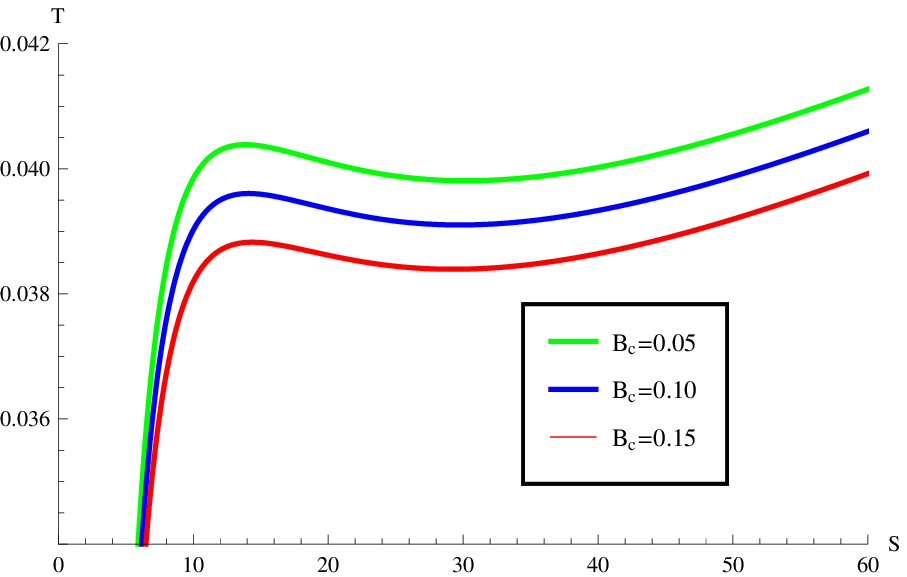}
\includegraphics[width=8.1cm,keepaspectratio]{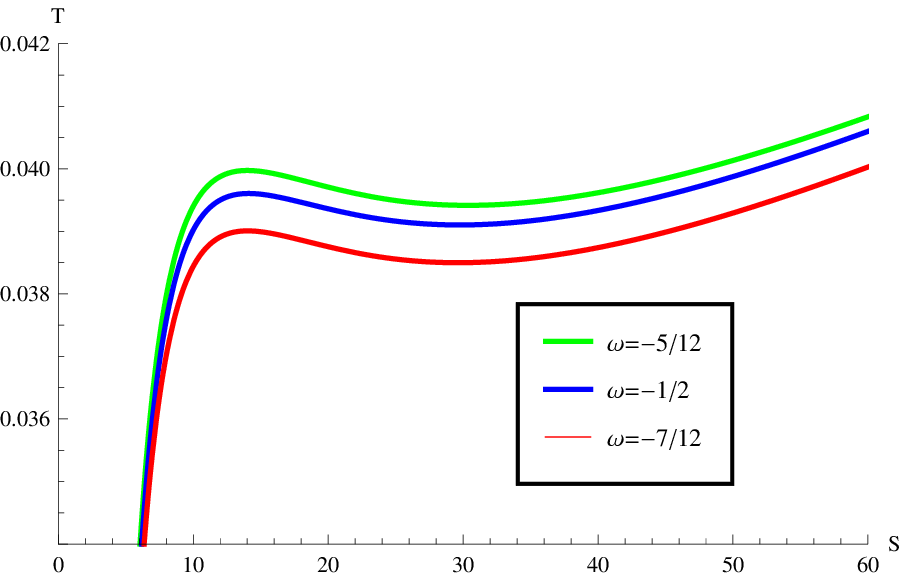}
\hspace{0.5cm}}\\
\captionsetup{font={scriptsize}}
\caption{(Color online) The simulated phase transition and the boundary of a two-phase
coexistence on the base of isobaric in the $T-S$ diagram for charged anti-de Sitter black holes.
with differential $B_c $ and $\omega $. The left fig. $Q = 1,\omega = - 1 / 2$, the right fig.,$Q = 1,B_c = 0.1$ \label{PVR2}}
\end{figure}

From Fig.(\ref{PVR2}), We find that at the same black hole pressure,
the temperature decreases, and the region of coexistence increases of the transition point increases as
the value of $B_c$ increases, the temperature decreases,
and the region of coexistence increases. While the temperature decreases and the
coexistence zone of the two phases increases with the decrease of
the value of $\omega$.

According to Eqs. (\ref{eq231}) and (\ref{eq17}), if the temperature $T_0$ and $\omega$ is given,
the location $r_2 $ or $r_1$ of the phase transition in charged AdSQ black hole
is related to the charged $Q$ and $\alpha$. Thus, the phase transition in charged AdSQ black
hole at a given temperature of $T_0$ depends on the electric potential $\phi _2$
and quintessence potential $\Theta _2$ at
the event horizon of the black hole, or $\phi_1$ and $\Theta _1$. It is not simply related to the event
horizon of the black hole $r_2$ or $r_1$.

For constant $Q$ and $\alpha$, at a given temperature $T_0$$(T_0<T_c)$, when the
radius of the event horizon of the black hole is $r_+ <r_1$, charged AdSQ black hole
corresponds to the liquid phase of the van der Waals system. When the event horizon
radius of the black hole is $r_+>r_2$, charged AdSQ black hole corresponds
to the vapor phase of the van der Waals system. When the event horizon radius
of the black hole is $r_1 \leq r_+\leq r_2$, charged AdSQ black hole corresponds
to the vapor liquid coexistence region of the van der Waals system.
Thus, according to the electric potential of three different
states, we get charged AdSQ black hole at a given temperature of $T < T_c$, the same pressure $P_0$,
the same charge $Q$ and $\alpha$, corresponding
to three different states, $\phi$ and $\Theta$ are respectively called high potential and high quintessence
potential phase (referred to as high coefficient phase), middle potential and middle quintessence
potential phase (referred to as medium coefficient phase) and low potential and low quintessence
potential phase (referred to as low coefficient phase). They, respectively, correspond to the liquid phase, liquid-vapor coexistence phase, and vapor phase of a van der Waals system.

\section{the latent heat of phase transition of charged AdSQ black hole}
\label{laten}
Due to the lack of complete knowledge of the chemical thermal potential of the ordinary
thermodynamic systems, the phase diagram, which is the $P-T$ curve when the two
phases $\alpha$ and $\beta$ are in equilibrium, is determined directly by experiments. The slope of the $P-T$
curve is determined by the Clapeyron equation
\begin{equation}
\label{eq31}
\frac{dP}{dT} = \frac{L}{T(v^\beta - v^\alpha )}
\end{equation}
where $L=T(s^\beta-s^\alpha )$, $s^\alpha$ and $s^\beta$ is the molar entropy of phase $\alpha$ and $\beta$,
$v^\alpha$ and $v^\beta$ is the molar volume of phase $\alpha$ and $\beta$.
For the general thermodynamic system, Clapeyron equation is in good agreement with
the experimental results, which provides a direct experimental verification for the correctness of thermodynamics.

For the charged AdSQ black hole thermodynamic system, substituting Eq. (\ref{eq17}) into Eq. (\ref{eq23})
to get the latent heat $L$ of the phase transition of the black hole
\[
L = \pi T(1 - x^2)r_2^2
\]
\begin{equation}
\label{eq32}
 = \pi (1 - x^2)r_2^2 \left[ {\frac{(1 + x)}{\pi r_2 x^2f_1 (x)} +
\frac{3\alpha }{4\pi xr_2^{3\omega + 2} }\left( {\frac{(1 + x)(1 + x^2)f_2
(x,\omega )}{1 + 4x + x^2} + \frac{\omega (1 - x^{3\omega + 3})}{x^{3\omega
+ 1}(1 - x)}} \right)} \right].
\end{equation}
Given $Q$, $\omega$ and $B_c $, by the Eqs. (\ref {eq17}) and (\ref {eq32}), the latent heat
of phase transition is a function of $x$. When $Q= 1$, the effects of different $\omega$
and $B_c$ on the latent heat of phase transition are shown in Tables (\ref{tab1}) and (\ref{tab2}).
\begin{table}[htbp]
\centering\caption{the latent heat of phase transition for different $\omega$ with $B_c=0.10$}
\begin{tabular}
{|p{40pt}<{\centering}|p{65pt}<{\centering}|p{106pt}<{\centering}|p{106pt}<{\centering}|p{106pt}<{\centering}|}
\hline
&
$B_c = 0.10$&
$\omega = - 7 / 12$&
$\omega = - 1 / 2$&
$\omega = - 5 / 12$ \\
\hline
\raisebox{-3.00ex}[0cm][0cm]{$L$}&
$x = 0.4$&
\textsf{0.5557}&
\textsf{0.5561}&
\textsf{0.5572} \\
\cline{2-5}
 &
$x = 0.5$&
\textsf{0.3921}&
\textsf{0.3923}&
\textsf{0.3930} \\
\cline{2-5}
 &
$x = 0.6$&
\textsf{0.2765}&
\textsf{0.2767}&
\textsf{0.2771} \\
\hline
\end{tabular}
\label{tab1}
\end{table}

\begin{table}[htbp]
\centering\caption{the latent heat of phase transition for different $B_c$ with $\omega=-1/2$}
\begin{tabular}
{|p{40pt}<{\centering}|p{65pt}<{\centering}|p{106pt}<{\centering}|p{106pt}<{\centering}|p{106pt}<{\centering}|}
\hline
&
$\omega = - 1 / 2$&
$B_c = 0.05$&
$B_c = 0.10$&
$B_c = 0.15$ \\
\hline
\raisebox{-3.00ex}[0cm][0cm]{$L$}&
$x = 0.4$&
\textsf{0.5597}&
\textsf{0.5561}&
\textsf{0.5526} \\
\cline{2-5}
 &
$x = 0.5$&
\textsf{0.3948}&
\textsf{0.3923}&
\textsf{0.3899} \\
\cline{2-5}
 &
$x = 0.6$&
\textsf{0.2784}&
\textsf{0.2767}&
\textsf{0.2750} \\
\hline
\end{tabular}
\label{tab2}
\end{table}

From the Table.(\ref{tab1}) and (\ref{tab2}), at the same value of $B_c $, the latent heat of phase transition
increases with the value of $\omega$ increase. At the same value of $\omega$, the latent
heat decreases with the increase of the value of $B_c$. As the value of $x$ increases,
the latent heat decreases. These Table.(\ref{tab1}) and (\ref{tab2}) show that at the same temperature and
pressure, the black hole from $r_1$ to $r_2$ through the two-phase coexistence zone needs
to absorb heat -- the latent heat of phase transition, which means that the black hole
molecules inside the black hole have different microstructure when they are in phase 1 and 2.

\section{the microcosmic explanation of phase transition for charged AdSQ black hole}
\label{micro}
\begin{figure}[!htbp]
\center{\includegraphics[width=8.1cm,keepaspectratio]{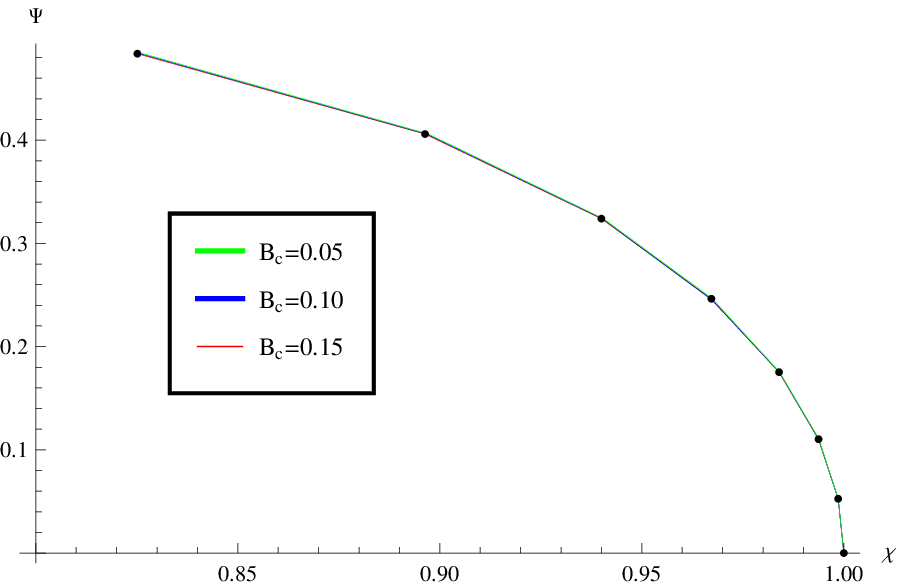}
\includegraphics[width=8.1cm,keepaspectratio]{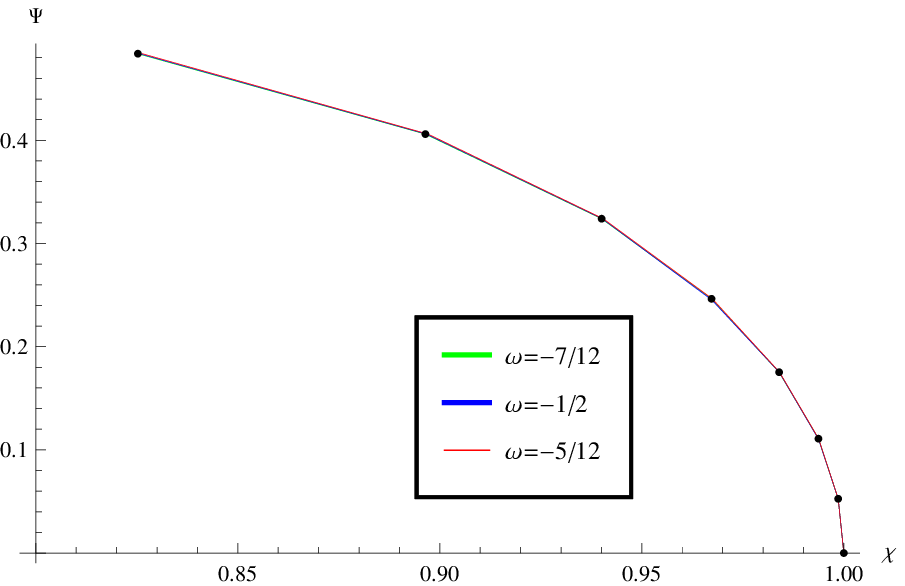}
\hspace{0.5cm}}\\
\captionsetup{font={scriptsize}}
\caption{(Color online) The $\Psi (T) - \chi $ diagram for charged anti-de Sitter black holes with differential $B_c
$ and $\omega $. The left fig. $Q = 1,\omega=-1/2$, the right fig.$Q=1,B_c=0.1$
\label{PVR3}}
\end{figure}

According to Eqs.(\ref{eq17}) and (\ref{eq231}), given $T_0$, $B_c$ and $\omega$,
when the black hole system phase transition occurs, the molecular potential
$\phi = \frac {Q} {r_ +} $ and quintessence potential $\Theta = - \frac{B_c r_c^{3\omega + 1} }{2\omega \pi r_+ ^{3\omega + 1} }=\frac{3\alpha }{r_ + ^{3\omega + 1}
}$ has a mutation, which reflects the inconsistent microstructure of the black hole molecules in different phases.
When the temperature is fixed, the electric potential and quintessence potential of the two-phase are respectively determined
\begin{equation}
\label{eq33}
\phi _2 = \frac{Q}{r_2 },
\quad
\Theta _2 = \frac{3\alpha }{r_2^{3\omega + 1} },
\quad
\phi _1 = \frac{Q}{r_1 } = \frac{Q}{xr_2 },
\quad
\Theta _1 = \frac{3\alpha }{x^{3\omega + 1}r_2^{3\omega + 1} },
\end{equation}
where $x$ is given by the Eq. (\ref{eq231}). When the temperature $T_0 \leq T_c $, $0 < x \leq 1$, one take the
order parameter
\begin{equation}
\label{eq34}
\Psi (T) = \frac{\phi _1 - \phi _2 }{\phi _c } + \frac{\Theta _1 -
\Theta _2 }{\Theta _c } = - \frac{\Delta \phi }{\phi _c } -
\frac{\Delta \Theta }{\Theta _c } = \frac{r_c (1 - x)}{r_2 x} +
\frac{r_c^{3\omega + 1} (1 - x^{3\omega + 1})}{r_2^{3\omega + 1} x^{3\omega
+ 1}}.
\end{equation}
When $\chi$, $B_c$, $\omega$, and $Q$ are given, substituting Eq.(\ref{eq17})
into Eq.(\ref{eq231}), one can obtain the solution to $x$ and $r_2$ for the corresponding temperature.
Substituting $\chi$, $x$, $r_2$, and $r_c$ into Eq. (\ref{eq34}), we get $\Psi(T)$ in Fig.(\ref{PVR3}).
we find that the curves are coincident with differential $B_c$ and $\omega$. In other words, the quintessence parameters
$B_c$ and $\omega$ have little effect on order parameter.

Recently, the microstructure of phase transition of black holes has been studied, and the phase transition between
black holes of different sizes is caused by the different density of molecules number of black holes
with different sizes~\cite{Wei2015,Wei2019,Wei20192,Rupp2013,Miao2019,Miao1712}.
Considering the influence of charge and quintessence on phase transition,
we re-examined the physical mechanism of phase transition in charged AdSQ black hole.

The Landau theory of continuous phase transition is characterized by changes in the
degree of material order and the accompanying changes in the symmetry of matter.
Since the black hole has the characteristics of ordinary thermodynamics, whether
the phase transition of the black hole also has the symmetry change similar to
the phase transition of ordinary thermodynamics is a question that people generally pay attention to.
According to the above discussion, we believe that the phase transition in charged AdSQ black hole
is also a symmetry change. When the temperature of the black hole is lower than the
critical point temperature and the black hole molecule is at a high coefficient phase 1,
the black hole molecule inside the black hole is subjected to a strong electric potential $\phi$.
The black hole molecule has a certain orientation under the action of the strong electric
potential and the higher quintessence potential.
The black hole molecules are in a relatively ordered state with low symmetry. At the same temperature,
when the black hole molecule is in the low coefficient phase 2, the potential $\phi$ and $\Theta$ that
cause the orientation of the black hole molecule is weakened, and the order degree of the black
hole molecule is relatively low and the symmetry is relatively high. With the increase of temperature,
the thermal motion of black hole molecules tends to weaken the order orientation. But
the temperature is not too high, there are still some black hole molecules have
a certain orientation. When the temperature of the black hole is above
the critical temperature, the thermal motion of the black hole molecules increases
and the orientation of the black hole molecules tends to zero. The phase below the critical
temperature, has low symmetry, higher order and non-zero $\Psi(T)$. The phase above
the critical temperature has higher symmetry, lower order, and the order parameter $\Psi(T)$ is zero.
As the temperature decreases, the order parameter $\Psi (T)$ continuously changes from zero to non-zero at the critical point.

From the Landaus's opinion, the order parameter $\Psi(T)$ is a small amount near the critical point $T_c$,
and can expand Gibbs $G(T,\phi)$ near $T_c$ as a power of $\Psi(T)$.
Considering that the phase transition of the system is due to the change of order degree of black hole molecules inside the black hole, then the system is symmetric with respect to transformation $\phi\rightleftarrows -\phi$. When $\Psi(T)$ can be taken as the order
parameter and the thermodynamic quantity of the black hole $G(T,\phi)$ is expanded according
to the order parameter perturbation series, there is no odd power term of the
order parameter, but only even order term of the order parameter. So
\begin{equation}
\label{eq35}
G(T,\phi ) = G_0 (T) + \frac{1}{2}a(T)\Psi ^2 + \frac{1}{4}b(T)\Psi ^4 +
\cdots ,
\end{equation}
where $G_0(T)$ is the Gibbs function when $\Psi(T)=0$.
We determine the functional relationship between $\Psi(T)$ and temperature $T$
by using the minimum stable equilibrium Gibbs function at $T$, $P$ constant.
Note that the Gibbs function is written as $G(T,\phi)$, where $\Psi$ is
not a independent variable. According to the requirement that $G(T)$ has
minimum value under stable equilibrium state, $\Psi$ should have three solutions
\begin{equation}
\label{eq36}
\Psi = 0,
\quad
\Psi = \pm \sqrt { - \frac{a}{b}}.
\end{equation}
This solution $\Psi=0$ on behalf of the disordered state,
corresponding to $T > T_c$ temperature range at $a>0$.
this nonzero solution $\Psi=\pm\sqrt{-\frac{a}{b}}$ represents an
orderly state, corresponding to $T<T_c$ temperature range at $a<0$.

The order parameter continues to shift from zero to non-zero at $T_c$,
so there should be $a=0$ at $T=T_c$. we can just take near the critical point
\begin{equation}
\label{eq37}
a = a_0 \left( {\frac{T - T_c }{T_c }} \right) = a_0 t,
\quad
a_0 > 0,
\end{equation}
Because $\Psi=\pm \sqrt{-\frac{a}{b}}$ is real. When $T<T_c$, $a<0$,
so we give the constant $b>0$. From this we can get
\[
\Psi = 0,
\quad
t > 0
\]
\begin{equation}
\label{eq38}
\Psi = \pm \left( {\frac{a_0 }{b}} \right)^{1 / 2}( - t)^{1 / 2},
\quad
t < 0.
\end{equation}
Various ferromagnetic systems have been found to have the following cooperative experimental laws in the critical point field:
\begin{itemize}
\item When $t=\frac{t-t_c}{T_c} \to -0 $, the spontaneous magnetization follows the following rules
\begin{equation}
\label{eq39}
M \propto ( - t)^\beta ,
\quad
t \to - 0.
\end{equation}

\item The zero field magnetization of various ferromagnetic substances $\chi = \left( {\frac{\partial M}{\partial H}}
\right)_T $ is divergent at $t \to \pm 0$. The change rule of $\chi$ with $t$ is
\begin{equation}
\label{eq40}
\chi \propto t^{ - \gamma },
\quad
t \to + 0;
\quad
\chi \propto ( - t)^{ - \gamma '},
\quad
t \to - 0.
\end{equation}

\item At $t=0$ and very weak magnetic field, the relationship between the magnetization of $M$ and the external magnetic field of $H$ is
\begin{equation}
\label{eq41}
M \propto H^{1 / \delta }.
\end{equation}

\item At $t\to\pm 0$, the zero field specific heat capacity of ferromagnetic material $c_H(H = 0)$ follows the following law
\begin{equation}
\label{eq42}
c_H \propto t^{ - \bar {\alpha }},
\quad
t > 0;
\quad
c_H \propto ( - t)^{ - \bar {\alpha }'},
\quad
t < 0.
\end{equation}
\end{itemize}
The dependence relation $\Psi$ to $t$ from Eq.(\ref{eq38}) is the same
as Eq. (\ref{eq39}), the critical index of $\beta=1/2$. According to the method
of the literature~\cite{Guo}, We can obtain the critical exponent
$\bar{\alpha}=\bar{\alpha }'=0$, $\gamma=\gamma'=1$, $\delta=3$,
and the entropy of charged AdSQ black hole near the critical point.

The entropy of disordered phase is $S=S_0$, and the entropy of ordered phase is
\begin{equation}
\label{eq43}
S=S_0+\frac{a_0^2 t}{2bT_c}.
\end{equation}
When $t=0$, the entropy of the ordered phase is equal to the entropy of
the disordered phase. It means that is, the entropy of the black hole is
continuous at the critical point.

\section{Thermdynamic geometry of charged AdSQ black hole}
\label{geo}

From the discussion in Sec.\ref{micro}, we find that Eq.(\ref{eq35}) contains parameters of $a$ and $b$
related to specific system characteristics, but the critical exponent given above is irrelevant to $a$ and $b$.
And we know that there are similar problems in the general thermodynamic system. The radical cause is that
when we discuss the continuous phase transition, we do not consider the strong fluctuation of order parameters
in the critical point domain. Fortunately, the famous Ruppeiner geometry comes from the thermodynamic fluctuation
theory. The singularity of curvature scalar of corresponding thermodynamic geometry is studied to reveal the phase transition of black hole~\cite{Ruppeiner2008,Ruppeiner1995}.
Therefore, we can reveal the microstructure of the black hole molecules by studying the Ruppeiner geometry. We take ($S,P)$ fluctuation while with fixed $Q$, $B_c$ and $\omega$. One can obtain the Ruppeiner curvature scalar(Ricci scalar) as
\[
R=\left((8 P S+1) S^{\frac{3 (\omega +1)}{2}}-\pi  Q^2 S^{\frac{1}{2} (3 \omega +1)}+3 \alpha  S \pi ^{\frac{1}{2} (3 \omega +1)} \omega \right)
\]
\[
\times \left(-2 \pi  Q^2 (16 P S+3) S^{3 \omega +1}+2 (8 P S+1) S^{3 \omega +2}+3 \alpha  \pi ^{\frac{1}{2} (3 \omega +1)} \omega  S^{\frac{3 (\omega +1)}{2}} (24 P S (\omega +1)+3 \omega +5) \right.
\]
\[
\left. +4 \pi ^2 Q^4 S^{3 \omega }-3 \alpha  Q^2 \pi ^{\frac{3 (\omega +1)}{2}} \omega  (3 \omega +7) S^{\frac{3 (\omega +1)}{2}}+27 \alpha ^2 S \pi ^{3 \omega +1} \omega ^2 (\omega +1)\right)
\]
\begin{equation}
/2 S^{3/2} (S^{3 \Omega/2} (-\pi Q^2 + S + 8 P S^2) +3 \alpha \pi^{(1 + 3 \Omega)/2 } \sqrt{S} \Omega)^3
\label{eq44}
\end{equation}
At the same temperature $R$ is divided into two parts, one representative
$\phi _2 = \frac{Q}{r_2 }$, $\Theta _2= - \frac{1}{2r_2^{3\omega } }$,
another representative $\phi _1 = \frac{Q}{r_1 } =
\frac{Q}{xr_2 }$, $\Theta _1 = - \frac{1}{2x^{3\omega }r_2^{3\omega }
}$. Substitute Eqs.(\ref{eq14}) and (\ref{eq16}) into Eq.(\ref{eq44}) and get
the value of $R$ with fixed $B_c$, $\omega$ and $Q$.

\begin{table}[htbp]
\centering\caption{the Ruppeiner curvature scalar for different $\omega$ with $B_c=0.10$}
\begin{tabular}
{|p{47pt}<{\centering}|p{46pt}<{\centering}|p{46pt}<{\centering}|p{52pt}<{\centering}|p{46pt}<{\centering}|p{46pt}<{\centering}|p{52pt}<{\centering}|p{46pt}<{\centering}|p{46pt}<{\centering}|p{52pt}<{\centering}|}
\hline
\raisebox{-1.50ex}[0cm][0cm]{$B_c = 0.10$}&
\multicolumn{3}{|p{142pt}<{\centering}|}{$\omega = - 7 / 12$} &
\multicolumn{3}{|p{142pt}<{\centering}|}{$\omega = - 1 / 2$} &
\multicolumn{3}{|p{142pt}<{\centering}|}{$\omega = - 5 / 12$}  \\
\cline{2-10}
 &
$R(r_1 )$\textsf{}&
$R(r_2 )$\textsf{}&
$R(r_1 )-
R(r_2 )$\textsf{}&
$R(r_1 )$\textsf{}&
$R(r_2 )$\textsf{}&
$R(r_1 )-
R(r_2 )$\textsf{}&
$R(r_1 )$\textsf{}&
$R(r_2 )$\textsf{}&
$R(r_1 )-
R(r_2 )$\textsf{} \\
\hline
$x = 0.3$&
\textsf{0.008920}&
\textsf{0.005250}&
\textsf{0.003670}&
\textsf{0.010071}&
\textsf{0.005520}&
\textsf{0.004550}&
\textsf{0.010580}&
\textsf{0.005439}\textsf{}&
\textsf{0.005140} \\
\hline
$x = 0.4$&
\textsf{0.016951}&
\textsf{0.008347}&
\textsf{0.008604}&
\textsf{0.018251}&
\textsf{0.008790}&
\textsf{0.009461}&
\textsf{0.018689}&
\textsf{0.008688}&
\textsf{0.010001} \\
\hline
$x = 0.5$&
\textsf{0.021818}&
\textsf{0.011697}&
\textsf{0.010120}&
\textsf{0.023264}&
\textsf{0.012345}&
\textsf{0.010918}&
\textsf{0.024008}&
\textsf{0.012599}&
\textsf{0.011408} \\
\hline
$x = 0.6$&
\textsf{0.024792}&
\textsf{0.015153}&
\textsf{0.009639}&
\textsf{0.026376}&
\textsf{0.016034}&
\textsf{0.010341}&
\textsf{0.026677}&
\textsf{0.015927}&
\textsf{0.010750} \\
\hline
$x = 0.7$&
\textsf{0.026583}&
\textsf{0.018608}&
\textsf{0.007975}&
\textsf{0.028294}&
\textsf{0.019747}&
\textsf{0.008546}&
\textsf{0.028530}&
\textsf{0.019657}&
\textsf{0.008872} \\
\hline
$x = 0.8$&
\textsf{0.027608}&
\textsf{0.021985}&
\textsf{0.005623}&
\textsf{0.029438}&
\textsf{0.023405}&
\textsf{0.006032}&
\textsf{0.029610}&
\textsf{0.023347}&
\textsf{0.006263} \\
\hline
$x = 0.9$&
\textsf{0.028127}&
\textsf{0.025230}&
\textsf{0.002897}&
\textsf{0.030067}&
\textsf{0.026952}&
\textsf{0.003115}&
\textsf{0.030178}&
\textsf{0.026941}&
\textsf{0.003236} \\
\hline
$x = 1$&
\textsf{0.028391}&
\textsf{0.028391}&
\textsf{0}&
\textsf{0.030361}&
\textsf{0.030360}&
\textsf{0}&
\textsf{0.030176}&
\textsf{0.030176}&
\textsf{0} \\
\hline
\end{tabular}
\label{tab3}
\end{table}

\begin{table}[htbp]
\centering\caption{the Ruppeiner curvature scalar for differential $B_c$ with $\omega=-1/2$}
\begin{tabular}
{|p{47pt}<{\centering}|p{46pt}<{\centering}|p{46pt}<{\centering}|p{52pt}<{\centering}|p{46pt}<{\centering}|p{46pt}<{\centering}|p{52pt}<{\centering}|p{46pt}<{\centering}|p{46pt}<{\centering}|p{52pt}<{\centering}|}
\hline
\raisebox{-1.50ex}[0cm][0cm]{$\omega = - 1 / 2$}&
\multicolumn{3}{|p{142pt}<{\centering}}{$B_c = 0.05$} &
\multicolumn{3}{|p{142pt}<{\centering}}{$B_c = 0.10$} &
\multicolumn{3}{|p{142pt}<{\centering}|}{$B_c = 0.15$}  \\
\cline{2-10}
 &
$R(r_1 )$\textsf{}&
$R(r_2 )$\textsf{}&
$R(r_1 )-
R(r_2 )$\textsf{}&
$R(r_1 )$\textsf{}&
$R(r_2 )$\textsf{}&
$R(r_1 )-
R(r_2 )$\textsf{}&
$R(r_1 )$\textsf{}&
$R(r_2 )$\textsf{}&
$R(r_1 )-
R(r_2 )$\textsf{} \\
\hline
$x = 0.3$&
\textsf{0.008903}&
\textsf{0.005295}&
\textsf{0.003608}&
\textsf{0.010071}&
\textsf{0.005520}&
\textsf{0.004550}&
\textsf{0.011255}&
\textsf{0.005752}&
\textsf{0.005503} \\
\hline
$x = 0.4$&
\textsf{0.016951}&
\textsf{0.008409}&
\textsf{0.008531}&
\textsf{0.018251}&
\textsf{0.008790}&
\textsf{0.009461}&
\textsf{0.019582}&
\textsf{0.009180}&
\textsf{0.010402} \\
\hline
$x = 0.5$&
\textsf{0.021821}&
\textsf{0.011776}&
\textsf{0.010045}&
\textsf{0.023264}&
\textsf{0.012345}&
\textsf{0.010918}&
\textsf{0.024732}&
\textsf{0.012929}&
\textsf{0.011802} \\
\hline
$x = 0.6$&
\textsf{0.024813}&
\textsf{0.015245}&
\textsf{0.009568}&
\textsf{0.026376}&
\textsf{0.016034}&
\textsf{0.010341}&
\textsf{0.027969}&
\textsf{0.016844}&
\textsf{0.011124} \\
\hline
$x = 0.7$&
\textsf{0.026623}&
\textsf{0.018709}&
\textsf{0.007914}&
\textsf{0.028294}&
\textsf{0.019747}&
\textsf{0.008546}&
\textsf{0.030000}&
\textsf{0.020812}&
\textsf{0.009187} \\
\hline
$x = 0.8$&
\textsf{0.027670}&
\textsf{0.022091}&
\textsf{0.005578}&
\textsf{0.029438}&
\textsf{0.023405}&
\textsf{0.006032}&
\textsf{0.031246}&
\textsf{0.024753}&
\textsf{0.006493} \\
\hline
$x = 0.9$&
\textsf{0.028211}&
\textsf{0.025339}&
\textsf{0.002872}&
\textsf{0.030067}&
\textsf{0.026952}&
\textsf{0.003115}&
\textsf{0.031968}&
\textsf{0.028607}&
\textsf{0.003361} \\
\hline
$x = 1$&
\textsf{0.028418}&
\textsf{0.028418}&
\textsf{0}&
\textsf{0.030361}&
\textsf{0.03036}&
\textsf{0}&
\textsf{0.032355}&
\textsf{0.032355}&
\textsf{0} \\
\hline
\end{tabular}
\label{tab4}
\end{table}
From the Tables (\ref{tab3}) and (\ref{tab4}), two curves of $R$ will overlap at $x = 1$ with differential $B_c $ and $\omega$.
For anyon gas curvature scalar $R > $0 (or $R <0$) means that the average interaction between particles is repulsive or attractive, while the average interaction is zero as $R =0$~\cite{Miao2019,Miao1712,Miao1711,Miao1610,Altamirano2014,Mirza2008,Mirza2009}.
From the Tables (\ref{tab3}) and (\ref{tab4}), we can find that $0 < R(r_2 ) < R(r_1)$, and think that the average attraction between the black hole molecules
of phase 2 in black holes is less than phase 1. According to the relation Eq. (\ref{eq1}) between the number density of black hole molecules and the position of the event horizon, the number density of black hole molecules in phase 2 is less than that in phase 1. The number density of black hole molecules decreases and mutual attraction decreases, which satisfies Lenard-Jones' description of the interaction potential between particles. When the distance between particles increases, the interaction potential between particles decreases, that is, when the number density of particles decreases, the interaction potential decreases.

From the Tables (\ref{tab3}) and (\ref{tab4}), for constant $B_c $, $R(r_1)$ and $R(r_2)$  will
change as $\omega$ and $x$ change, $R(r_1 )-R(r_2 )$ will increase with $\omega$ increase.
When $\omega $ is fixed $R(r_1 )-R(r_2 )$ will increase with $B_c $ increase. Whatever 
the values of $B_c$ and $\omega$, $R(r_1 )-R(r_2 )$ starts to increase as $x$ increases 
until it reaches its maximum value at $x=0.5$, then $R(r_1 )-R(r_2 )$ starts to decrease, 
and finally $R(r_1 )-R(r_2 )=0$ when $x=1$. According to Eq.(\ref{eq8}), 
the change of $B_c$ and $\omega$ with the same temperature and pressure is
accompanied by the change of the radius of the event horizon of the black hole. Therefore, the change
of $B_c$ and $\omega$ makes the number of black hole molecules density
change with the temperature and pressure fixed. So, we think that the charge
$Q$, $B_c $ and $\omega$ play two roles in phase transition of black hole, on the one hand,
they can change the order degree of black hole molecules, on the other hand, they can change the number
density of black hole molecules. These two aspects are the main reasons for the phase transition of thermodynamic system.

\section{Discussion and Summary}
\label{con}
The investigation of the thermal properties and the internal microstructure of black holes has always been one of the topics of interest to theoretical physicists.
Although the precise statistics description of the thermodynamic states corresponding to black hole is unclear, the black hole thermodynamics as well as the critical phenomenon is still a popular topic. Since the discovery of the accelerated expansion of the universe, dark energy becomes one of the topics that many physicists are interested in. Among the many dark energy models, quintessence is very influential.

In this paper, we choose different conjugate variables and use Maxwell's equal-area law to
study the phase transition of charged AdSQ black hole. We find that the phase transition points
which obtained by choosing the conjugate variable $(P,V)$ and $(T,S)$, respectively, are same for the same temperature.
It was found that the charged AdSQ black hole had a phase transition similar to that of vdW system when the temperature
was below the critical temperature, and the charged AdSQ black hole had a phase transition similar to that of
vdWs system under the same temperature and pressure. From Eq. (\ref{eq17}), whether a black hole experiences
a first order phase transition depends on the electric potential at the event horizon $\phi
= \frac{Q}{r_ + }$ and quintessence potential $\Theta=-\frac{B_c r_c^{3\omega + 1}
}{2\omega \pi r_ + ^{3\omega + 1} } = \frac{3\alpha }{r_ + ^{3\omega + 1}
}$, not just the size of the black hole. It shows that the microstructure of black hole molecules will change under the action of electric potential and quintessence potential. Through the analysis in Section (\ref{laten}), we obtained that the charged AdSQ black hole also had the phase transition latent heat similar to the vdW system in the phase transition of the first order.

From the microscopic point of view, fluid is made up of fluid molecules, so it can be speculated that black hole is also made up of black hole molecules, and black hole molecules themselves carry the microscopic degree of freedom of black hole entropy~\cite{Wei2015,Wei2019,Wei20192,Miao2019,Miao1712,Miao1711,Miao1610,Guo}. Considering that the phase transition of charged AdSQ black hole at a certain temperature and pressure is not simply a phase transition from a large black hole to a small black hole, but is determined by the electric potential and quintessence potential at the event horizon. The charge $Q$ of black hole, dark energy state parameter $\omega$ and the normalized factor $\alpha$ related to the dark energy density play a key role in the phase transition, so we introduced the order parameter of the charged AdSQ black hole. In Section. (\ref{micro}), Landau continuous phase transition theory is used to analyze the critical phenomenon of black hole, and the critical index is obtained. The Section.(\ref{geo}), we reveals the microstructure of the black hole molecules by studying the geometry Ruppeiner of the black hole system. According to the standard curvature $R$ sign statistical interpretation,curvature scalar  $R>0$ (or $R <0$) means that the average interaction between particles is repulsive or attractive, while the average interaction is zero as $R =0$.
According to the given conclusion, the average black hole molecules in the interior of charged AdSQ black hole are attracted to each other. We have analyzed that the influence of dark energy state parameter $\omega$ and dark energy density related normalized factor $\alpha$ on phase transition and latent heat of phase transition.

This work revealed that the microstructure of charged AdSQ black hole was similar to that of vdW system.These conclusions are helpful for people to explore the microstructure of black holes. The study of black hole microstructure is the basis of quantum gravity theory and the bridge for people to understand the relationship between quantum mechanics and gravity theory. In particular, the in-depth study of black hole microstructure will help to understand the basic properties of black hole gravity, and is of great significance to the establishment of quantum gravity.

\section*{Acknowledgements}
We would like to thank Prof. Zong-Hong Zhu and Meng-Sen Ma for their indispensable discussions and comments. This work was supported by the Young Scientists Fund of the National Natural Science Foundation of China (Grant No.11205097), in part by the National Natural Science Foundation of China (Grant No.11475108), Supported by Program for the Innovative Talents of Higher Learning Institutions of Shanxi, the Natural Science Foundation of Shanxi Province,China(Grant No.201601D102004).

\end{document}